\documentclass[apj]{emulateapj}

\usepackage{graphicx,times}
\usepackage{subfigure}
\newcommand{\be}{\begin{equation}}
\usepackage{threeparttable}
\usepackage{booktabs}
\newcommand{\ee}{\end{equation}}
\newcommand{\bea}{\begin{eqnarray}}
\newcommand{\eea}{\end{eqnarray}}

\usepackage{enumerate}
\usepackage{amsmath}
\usepackage{cases}
\usepackage{longtable}
\usepackage{hyperref}
\usepackage{epstopdf}
\usepackage{amsmath,bm}
\usepackage{amssymb}
\usepackage{natbib}
\usepackage{morefloats}
\usepackage{multirow}
\usepackage{array}
\usepackage{verbatim}

\begin{document}

\title{Nonuniversal interstellar density spectra probed by pulsars}

\author{Siyao Xu\altaffilmark{1} and Bing Zhang \altaffilmark{2}}

\altaffiltext{1}{Institute for Advanced Study, 1 Einstein Drive, Princeton, NJ 08540, USA;
Hubble Fellow, sxu@ias.edu}
\altaffiltext{2}{Department of Physics and Astronomy, University of Nevada Las Vegas, NV 89154, USA; zhang@physics.unlv.edu}

\begin{abstract}

The Galactic interstellar turbulence affects the density distribution and star formation. 
We introduce a new method of measuring interstellar turbulent density spectra by using the dispersion measures (DMs) of a large sample of pulsars.
Without the need of invoking multiple tracers, we obtain nonuniversal density spectra 
in the multi-phase interstellar medium over different ranges of length scales. 
By comparing the analytical structure function of DMs with the observationally measured one in different areas of sky, 
we find a shallow density spectrum arising from the supersonic turbulence in cold interstellar phases, 
and a Kolmogorov-like density spectrum in the diffuse warm ionized medium (WIM). 
Both spectra extend up to hundreds of pc.  
On larger scales, we for the first time identify a steep density spectrum in the diffuse WIM extending 
up to several kpc. 
Our results show that the DMs of pulsars can
provide unique new information on the interstellar turbulence.

\end{abstract}


\section{Introduction}

Turbulence is ubiquitous in astrophysical media and occurs over a vast range of length scales. 
Observations show that the turbulent spectrum of electron density in the Galactic interstellar medium (ISM) extends from 
$10^6$ to $10^{18}$ m 
\citep{Armstrong95,CheL10,Lee19}.
In the intergalactic medium, a turbulent spectrum of electron density up to $100$ Mpc 
has recently been inferred from the observations of fast radio bursts
\citep{XZ20}. 
Turbulence participates and plays an essential role 
in a variety of fundamental astrophysical processes
\citep{Brad13}.

The turbulence in the Galactic ISM and its astrophysical implications have been extensively studied 
\citep{ElmegreenScalo,Sca04}.
As one of its effects, turbulence shapes the density distribution in different interstellar phases and 
significantly influences star formation
\citep{MacL04,Mckee_Ostriker2007}.
This is demonstrated by numerous studies on density statistics
(e.g., \citealt{LG01,Pad20,BLC05,KL07,Burk09,Fed15,Kri17,XJL19}),
which are more easily accessible to observations compared with the statistics of, e.g., velocity and magnetic field. 
The power spectra of density fluctuations is frequently used as a diagnostic of turbulence properties. 
For instance, 
the spectral slope is dependent on the sonic Mach number of turbulence
\citep{KL07}.
The sonic Mach number varies in different interstellar phases 
\citep{Zuc74,Hei03,Gae11}
and is believed to be an important parameter affecting the star formation rate 
\citep{FedK12}.
The small inner scale of a shallow density spectrum, 
where turbulent energy is dissipated, 
corresponds to the correlation length of density structures, 
while the large outer scale of a steep density spectrum, 
where turbulent energy is injected, 
indicates the driving scale of turbulence 
\citep{LP04,LP06}.

The density spectra in different interstellar phases can be directly measured by using 
spatially continuous emission with the corresponding gas tracers and dust
\citep{Laz09rev,CheL10,HF12,Pin18}.
From the perspective of pulsar observations, 
a density spectrum over a wide range of length scales has been suggested based on interstellar scintillation and scattering 
for several decades
\citep{Le76,Arm81,Armstrong95}.
In particular, a shallow density spectrum in cold interstellar phases was identified from temporal broadening measurements 
of pulsars by 
\citet{XuZ17}.

Different from earlier approaches to obtain interstellar density spectra, 
here we employ the dispersion measures (DMs) of pulsars and extract the underlying interstellar density spectra from the 
structure functions (SFs) of their DMs. 
The relation between the SF of projected quantities and 
the 3D power-law spectrum of their fluctuations was established by 
\citet{LP16} (hereafter LP16),
which depends on the shallowness of the turbulent power-law spectrum and the 
thickness of the observed turbulent volume. 
The LP16 analytical approach was proposed for studying the  
Faraday rotation and synchrotron statistics of a spatially continuous
synchrotron-emitting medium. 
It has been extended to point sources, including molecular cloud cores, fast radio bursts, radio sources, 
and a variety of observables, e.g., DMs, emission measures, velocities
\citep{XuZ16,Xu20,XZ20}.
In this work, we apply the LP16 approach to the DMs of pulsars.
By using many lines of sight through the ISM toward a large number of pulsars, 
we are able to sample the turbulence in both warm and cold interstellar phases 
over a broad range of length scales, with no need to invoke multiple tracers. 
In Section 2, we present the formalism of the DM SF in different scenarios. 
In Section 3, we analyze the observations of pulsars with our statistical approach to
obtain the interstellar density spectra. 
The discussion and conclusions are given in Section 4.

\section{SF of DMs}
\label{sec: sfdm}

We describe the statistically homogeneous electron density $n_e$ as a sum of the mean density $n_{e0}$ and a density fluctuation 
$\delta n_e$, 
\begin{equation}\label{eq: assmfd}
   n_e = n_{e0} + \delta n_e,
\end{equation}
where $\delta n_e$ is a zero mean fluctuation. 
For electron density fluctuations induced by turbulence, 
the two-point correlation function (CF) takes the form (LP16),
\begin{equation}\label{eq: cfvz}
\begin{aligned}
      \xi(R,\Delta l) &=  \langle \delta n_e (\bm{X_1}, l_1) \delta n_e (\bm{X_2,},l_2)\rangle \\
    &  = \langle \delta n_e^2 \rangle   \frac{L_i^m}{L_i^m  +  (R^2 + \Delta l^2)^\frac{m}{2}} ,
\end{aligned}
\end{equation}
where $\bm{X}$ is the position on the sky plane, $l$ is the distance along the line of sight (LOS), 
$R= |\bm{X_1} - \bm{X_2}|$, $\Delta l = l_1 - l_2 $,
and $\langle ... \rangle$ means an ensemble average. 
The statistical properties of $\delta n_e$ are described by the correlation length $L_i$ and 
the power-law index $m$. 
The latter is related to the spectral index of turbulence $\alpha$ by 
\citep{LP06}, 
\begin{subnumcases}
{\alpha = \label{eq: malp}}
  m-N, ~~~~~~ \alpha_{3D} >-3,  ~~\text{shallow}, \label{eq: alaslw} \\
  -m-N, ~~~\alpha_{3D} <-3,  ~~\text{steep},
\end{subnumcases}
where $N$ is the dimensionality of space. 
A ``shallow" spectrum with the 3D spectral index $\alpha_{3D} >-3$
is dominated by small-scale density fluctuations, 
while a ``steep" spectrum with $\alpha_{3D} <-3$ is dominated by large-scale density fluctuation. 
In supersonic turbulence with local density enhancements due to shock compression, 
the density spectrum tends to be shallow
\citep{BLC05,KL07}.
In subsonic turbulence, the steep Kolmogorov density spectrum with $\alpha_{3D} = -11/3$ is usually expected
\citep{Armstrong95,CheL10}.

As a handy observable of pulsars, 
the dispersion measure DM$= \int n_e dl$ is the column density of free electrons between the observer and the source. 
The SF of the DMs of a sample of point sources in a turbulent medium is
\begin{align}
     D(R,l_1,l_2) & =  \langle [\text{DM} (\bm{X_1},l_1) - \text{DM} (\bm{X_2},l_2)]^2 \rangle  \nonumber\\
                           & = \Big\langle \Big[ \int_0^{l_1} dl n_e (\bm{X_1},l) - \int_0^{l_2} dl n_e (\bm{X_2}, l) \Big]^2 \Big\rangle \nonumber\\
                           & = \Big\langle \Big[ \int_0^{l_1} dl \delta n_e (\bm{X_1},l)  - \int_0^{l_2} dl \delta n_e (\bm{X_2}, l) \Big]^2 \Big\rangle \nonumber\\
                           &    ~~~~~~ + n_{e0}^2  (\Delta l)^2  \nonumber\\
                           & = D_{\delta n_e}(R,l_1,l_2) + n_{e0}^2  (\Delta l)^2. \label{eq: genllno}
\end{align}   
Here
\begin{equation}
\begin{aligned}
   &  D_{\delta n_e}(R,l_1,l_2)  \\
     = & \Big\langle \Big[ \int_0^{l_1} dl \delta n_e (\bm{X_1},l)  - \int_0^{l_2} dl \delta n_e (\bm{X_2}, l) \Big]^2 \Big\rangle 
\end{aligned}
\end{equation}
is the SF of DM fluctuations, 
which contains the information about the statistical properties of turbulence. 
The second term in Eq. \eqref{eq: genllno} comes from the difference between the distances of a pair of point sources, 
and its value depends on $n_{e0}$.

The approximate expression of $D_{\delta n_e}(R,l_1,l_2)$ is (LP16)
\begin{equation}
    D_{\delta n_e}(R,l_1,l_2) \approx 2 D^+(R,l_+) + \frac{1}{2} \Lambda (\Delta l)^2 ,
\end{equation}
with
\begin{equation}\label{eq: dpluf}
\begin{aligned}
     D^+(R,l_+) & = 2  \langle \delta n_e^2 \rangle \int_0^{l_+} d\Delta l (l_+ -\Delta l)  \\
     & ~~~~~~ \Bigg[    \frac{L_i^m}{L_i^m  +  \Delta l^m} -    \frac{L_i^m}{L_i^m  +  (R^2 + \Delta l^2)^\frac{m}{2}}  \Bigg] ,
\end{aligned}
\end{equation}
and 
\begin{align}\label{eq: lamg}
     \Lambda &= \xi(0,l_+) - \xi(R,l_+) + 2\xi(R,0) \nonumber\\
                       &= \langle \delta n_e^2 \rangle  \Bigg[ \frac{L_i^m}{L_i^m  +    l_+^m}  - 
                              \frac{L_i^m}{L_i^m  +  (R^2 + l_+^2)^\frac{m}{2}} \nonumber\\
                       & ~~~~   + 2   \frac{L_i^m}{L_i^m  +  R^m } \Bigg ], 
\end{align}
where $l_+ = (l_1 + l_2)/2$.
We next consider the simplified expressions of $D_{\delta n_e}(R,l_1,l_2)$ in different cases.

\subsection{Case 1: $l_+ > L_i $ and a shallow density spectrum}

When $l_+ > L_i $, $D^+(R,l_+)$ in Eq. \eqref{eq: dpluf} has asymptotic expressions in different regimes (LP16),
\begin{subnumcases}
     { D^+(R,l_+) \approx   }
       2  \langle \delta n_e^2 \rangle L_i^{-m} l_+ R^{m+1}, ~~~~~~R<L_i,   \label{eq: shthi1}\\    
       2  \langle \delta n_e^2 \rangle L_i^m l_+ R^{-m+1} , L_i<R<l_+,    \label{eq: shthi2}\\     
       2  \langle \delta n_e^2 \rangle L_i^m l_+^{-m+2}, ~~~~~~~~~~~ R > l_+. \label{eq: shthi3}
\end{subnumcases}
Here only Eqs. \eqref{eq: shthi2} and \eqref{eq: shthi3} apply to the case of a shallow density spectrum, 
which has the inner scale as $L_i$. 
Approximately, $\Lambda$ in Eq. \eqref{eq: lamg} at $l_+ > L_i $ becomes 
\begin{subnumcases}
   {\Lambda \approx}
   2 \langle \delta n_e^2 \rangle, ~~~~~~ R < L_i, \label{eq: lc1atse} \\
   0, ~~~~~~~~~~~~~~~~ R> L_i, \label{eq: shtapp}
\end{subnumcases}
{where we adopt Eq. \eqref{eq: shtapp} for the inertial range of turbulence.}
Therefore, we have $D(R,l_1,l_2)$ as 
\begin{subnumcases}
     { D(R,l_1,l_2)  \approx   \label{eq: c1dtp}}
       4  \langle \delta n_e^2 \rangle L_i^m l_+ R^{-m+1} + n_{e0}^2  (\Delta l)^2, \nonumber \\
       ~~~~~~~~~~~~~~~~~~~~~~~~~~~~~L_i<R<l_+,\\     
       4  \langle \delta n_e^2 \rangle L_i^m l_+^{-m+2} + n_{e0}^2  (\Delta l)^2, \nonumber \\
       ~~~~~~~~~~~~~~~~~~~~~~~~~~~~~~~~~~~~~R > l_+.
\end{subnumcases}
When we consider a sample of point sources with different distances, 
$l_+$ and $\Delta l$ are in the ranges $[L_i + |\Delta l| /2 , L- |\Delta l|/2]$ and $[L_i-L,L-L_i]$, respectively. 
Here $L$ is the thickness of the turbulent volume sampled by the point sources. 
We replace $l_+$ and $(\Delta l)^2$ with their averages 
$\langle l_+\rangle  = (L+L_i)/2$ and $\langle (\Delta l)^2 \rangle = 1/3 (L-L_i)^2$ in Eq. \eqref{eq: c1dtp}
and obtain 
\begin{subnumcases}
         {D(R) \approx  \label{eq: c1apdts} }
       2  \langle \delta n_e^2 \rangle L_i^m (L+L_i) R^{-m+1} + \frac{1}{3}   n_{e0}^2 (L-L_i)^2 , \nonumber \\
       ~~~~~~~~~~~~~~~~~~~~~~~~~~~~~~~~~~~~~L_i \lesssim R \lesssim \frac{L+L_i}{2}, \label{eq: aptcstdr}\\     
       4  \langle \delta n_e^2 \rangle L_i^m \bigg(\frac{L+L_i}{2}\bigg)^{-m+2} + \frac{1}{3}  n_{e0}^2  (L-L_i)^2 , \nonumber \\
       ~~~~~~~~~~~~~~~~~~~~~~~~~~~~~~~~~~~~~~~~~~~~~~ R \gtrsim \frac{L+L_i}{2}.
\end{subnumcases}
Only when the first term dominates over the second term in Eq. \eqref{eq: aptcstdr}, can the scaling of turbulence be seen.  
{The scale where $D(R)$ saturates depends on $L$.}
We stress the approximate nature of Eq. \eqref{eq: c1apdts}, as it results from the average over a sample of point sources 
with different distances.

\subsection{{Case 2: $l_+ > L_i $ and a steep density spectrum}}

A steep density spectrum has the outer scale as $L_i$. Hence, only Eq. \eqref{eq: shthi1} is applicable for $D^+(R,l_+)$. 
$D^+(R,l_+)$ tends to saturate at $R=L_i$ and remain constant at a larger $R$. 
Likewise, we should only use Eq. \eqref{eq: lc1atse} for $\Lambda$.
Then $D(R,l_1,l_2)$ has the form,
\begin{subnumcases}
     { D(R,l_1,l_2) \approx   }
       4  \langle \delta n_e^2 \rangle L_i^{-m}   l_+  R^{m+1} \nonumber \\ 
       +\big( \langle \delta n_e^2 \rangle   + n_{e0}^2\big)  (\Delta l)^2 , 
       ~~~R<L_i,\\    
       4  \langle \delta n_e^2 \rangle L_i l_+ +\big( \langle \delta n_e^2 \rangle  + n_{e0}^2\big)  (\Delta l)^2 , \nonumber \\
       ~~~~~~~~~~~~~~~~~~~~~~~~~~~~~~~~~~~~~~~~~R>L_i .
\end{subnumcases}
Similar to Case 1, we take the averages of $l_+$ and $(\Delta l)^2$ and find, 
\begin{subnumcases}
     { D(R) \approx  \label{eq: ca2thess} }
       2  \langle \delta n_e^2 \rangle L_i^{-m} (L+L_i) R^{m+1}  \nonumber \\
        + \frac{1}{3} \big(\langle \delta n_e^2 \rangle + n_{e0}^2\big) (L-L_i)^2 ,  
       ~~~~~~~~~R\lesssim L_i, \label{eq: tsslora}\\    
       2  \langle \delta n_e^2 \rangle L_i (L+L_i) + \frac{1}{3} \big(\langle \delta n_e^2 \rangle + n_{e0}^2\big) (L-L_i)^2  , \nonumber \\
       ~~~~~~~~~~~~~~~~~~~~~~~~~~~~~~~~~~~~~~~~~~~~~~~~~~~~~~~~~R\gtrsim L_i.
\end{subnumcases}
We note that $D(R)$ in Eq. \eqref{eq: tsslora} has a steeper scaling with $R$
than that in Eq. \eqref{eq: aptcstdr}. 
{The scales where $D(R)$ saturates in the two cases are also different. }

\subsection{{Case 3: $l_+ < L_i $ and a steep density spectrum}}

When the distances of the sources are smaller than $L_i$, $D^+(R,l_+) $ has asymptotic expressions 
(LP16),
\begin{subnumcases}
     { D^+(R,l_+)  \approx  \label{eq: drtic}}
       2  \langle \delta n_e^2 \rangle L_i^{-m}l_+R^{m+1}, ~~~~~R<l_+,  \label{eq: thinsa1}\\   
       2  \langle \delta n_e^2 \rangle  L_i^{-m}l_+^2 R^m,~~ l_+<R<L_i,  \label{eq: thinsa2}\\      
       2 \langle \delta n_e^2 \rangle l_+^2, ~~~~~~~~~~~~~~~~~~~~~~~~~ R > L_i.  \label{eq: thinsa3}
\end{subnumcases}
$\Lambda$ can also be approximately simplified as, 
\begin{subnumcases}
   {\Lambda \approx}
   2 \langle \delta n_e^2 \rangle, ~~~~~~ R < L_i, \label{eq: instethin}\\
   \langle \delta n_e^2 \rangle, ~~~~~~~~ R> L_i.
\end{subnumcases}
The case with $l_+ < L_i $ is only applicable for a steep density spectrum. 
Accordingly, in the above expression we can only use Eq. \eqref{eq: instethin}, 
So there is
\begin{subnumcases}
     { D(R,l_1,l_2)  \approx  }
       4  \langle \delta n_e^2 \rangle L_i^{-m}l_+R^{m+1} \nonumber \\
       + \big( \langle \delta n_e^2 \rangle + n_{e0}^2 \big) (\Delta l)^2, 
     ~~~~~~~~  R<l_+,  \\   
        4 \langle \delta n_e^2 \rangle  L_i^{-m}l_+^2 R^m +  \big( \langle \delta n_e^2 \rangle + n_{e0}^2 \big) (\Delta l)^2, \nonumber \\
        ~~~~~~~~~~~~~~~~~~~~~~~~~~~~~~~~~~~~ l_+<R<L_i,  \\      
        4 \langle \delta n_e^2 \rangle l_+^2 + \big( \langle \delta n_e^2 \rangle + n_{e0}^2 \big) (\Delta l)^2,  \nonumber \\
       ~~~~~~~~~~~~~~~~~~~~~~~~~~~~~~~~~~~~~~~~~~~~~~ R > L_i. 
\end{subnumcases}
The steepening of the $D(R,l_1,l_2)$-$R$ relation at $R < l_+$ comes from the projection effect when the 
turbulent volume in the LOS direction is relatively thick (see also Case 2). 
This feature can be used to determine the LOS thickness of a turbulence layer in observations 
\citep{LP00,Elm01,Pad01}.

Based on the above result, we take the average of $l_+$ over the range $[|\Delta l| /2 , L- |\Delta l|/2]$ 
and the average of $(\Delta l)^2$ over the range $[-L, L]$,
i.e., $\langle l_+\rangle = L/2$ and $\langle (\Delta l)^2\rangle =1/3 L^2$, where $L< L_i$,
and find 
\begin{subnumcases}
     { D(R)  \approx  \label{eq: ca3sthth}}
       2  \langle \delta n_e^2 \rangle L_i^{-m}L R^{m+1} + \frac{1}{3} \big(\langle \delta n_e^2 \rangle + n_{e0}^2 \big) L^2, \nonumber \\
     ~~~~~~~~~~~~~~~~~~~~~~~~~~~~~~~~~~~~~~~~~~~~~~~~~~~  R\lesssim \frac{L}{2},  \\   
         \langle \delta n_e^2 \rangle  L_i^{-m} L^2 R^m + \frac{1}{3} \big(\langle \delta n_e^2 \rangle + n_{e0}^2 \big) L^2, \nonumber \\
        ~~~~~~~~~~~~~~~~~~~~~~~~~~~~~~~~~~~~~~~~~~ \frac{L}{2}\lesssim R \lesssim L_i,  \\      
         \langle \delta n_e^2 \rangle L^2 +  \frac{1}{3} \big(\langle \delta n_e^2 \rangle + n_{e0}^2 \big) L^2, ~ R \gtrsim L_i. 
\end{subnumcases}

The above approximate expressions of $D(R)$ in different cases can be applied to a realistic situation with unknown distances of many point sources.
Depending on the slope of the observationally measured SF and the LOS thickness of the turbulent medium relative to the 
transverse separation $R$ between lines of sight, the formula of $D(R)$ in the corresponding regime should be used.

\section{Interstellar density spectra}

\begin{figure*}[htbp]
\centering   
\subfigure[]{
   \includegraphics[width=9cm]{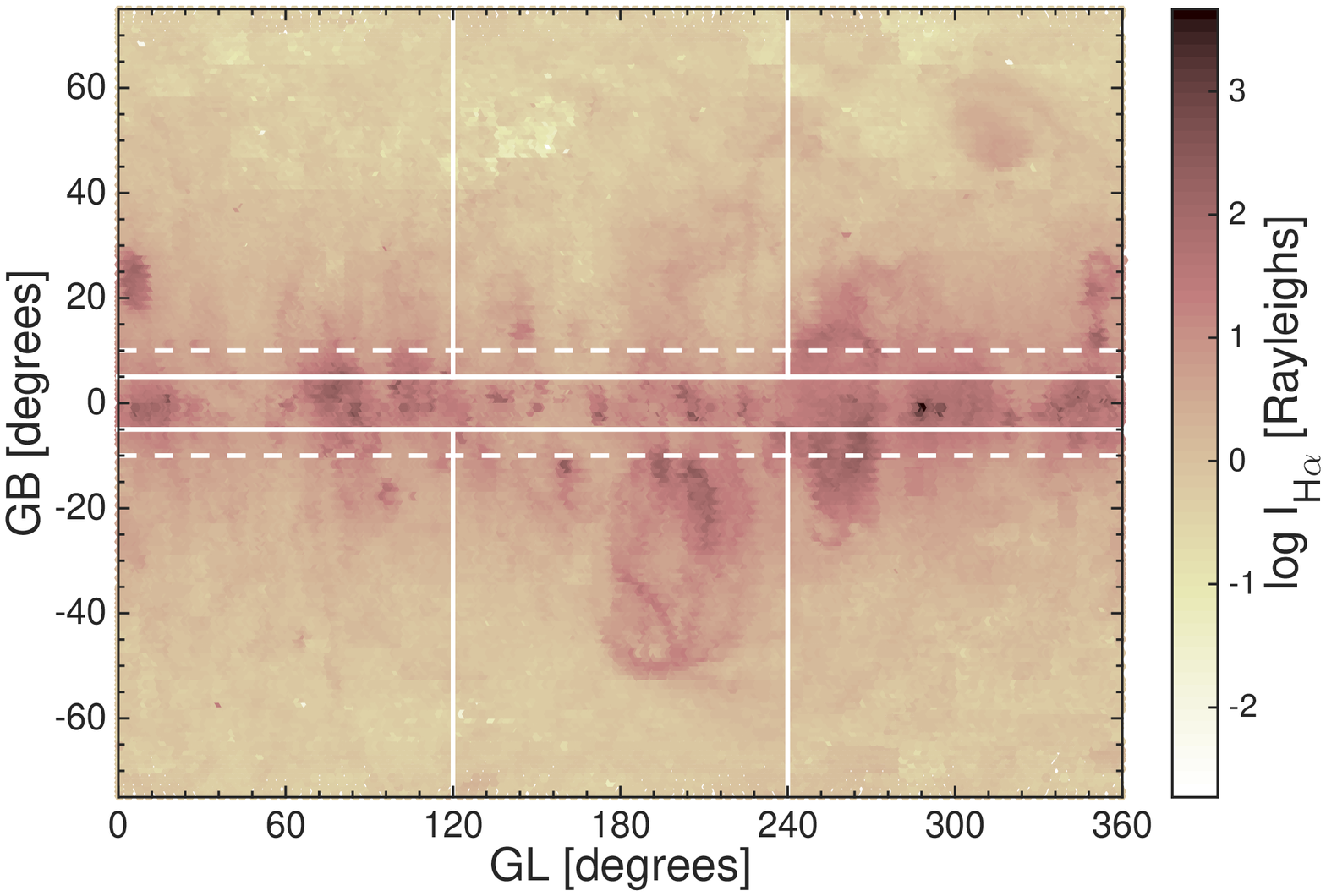}\label{fig: halp}}
\subfigure[]{
   \includegraphics[width=8.5cm]{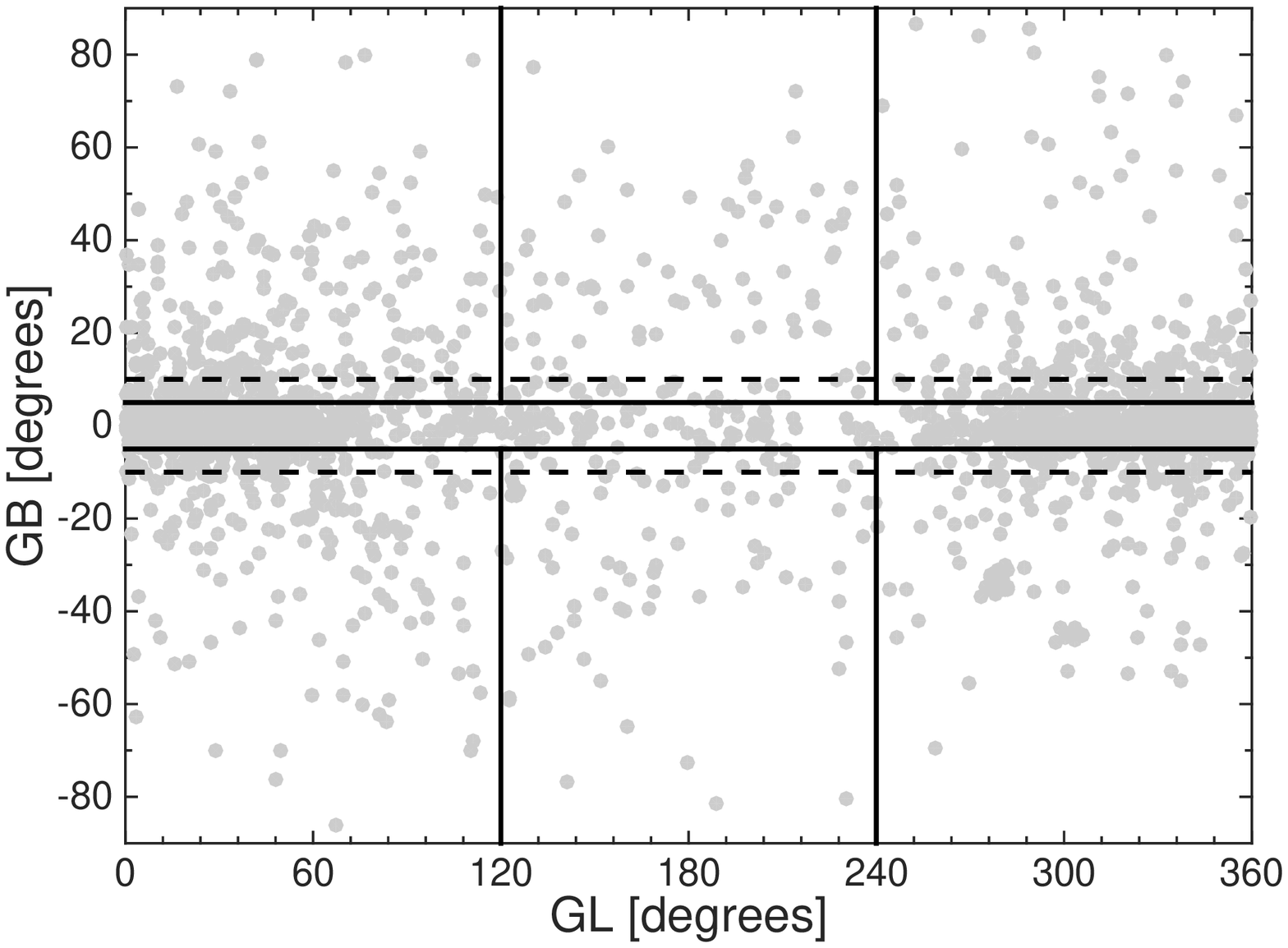}\label{fig: pulm}}
\caption{ (a) H$\alpha$ emission from WHAM sky survey. The color scale represents the 
logarithmic velocity-integrated intensity. 
(b) All sky map of pulsars from the ATNF Pulsar Catalogue. 
``GL" and ``GB" represent Galactic longitude and Galactic latitude. 
}
\label{fig: skymap}
\end{figure*}

The Wisconsin H-Alpha Mapper (WHAM) data 
\citep{Haf03,Haff10}
reveal the complex structure of the interstellar ionized hydrogen content 
(see Fig \ref{fig: halp}). 
The discrete bright clumps indicate H \uppercase\expandafter{\romannumeral2} regions and are associated with massive star formation 
in the Galactic disk. 
The diffuse H$\alpha$ emission arises from the warm ionized medium (WIM). 
The electron density fluctuations in the WIM obtained from the WHAM data were found to have a 
Kolmogorov power spectrum
\citep{CheL10},
as a remarkable extension of the Big Power Law in the sky earlier derived from interstellar scintillations and scattering
in the local ISM ($\lesssim 1$ kpc)
\citep{Armstrong95}.

The DMs of Galatic pulsars are the integrals of $n_e$ along many different paths through the ISM. 
Their SF can be used to study the fluctuations in DMs on different length scales in the multi-phase ISM, 
from which we can extract the statistical properties of $\delta n_e$ induced by the interstellar turbulence. 
For the SF analysis, we take 2719 pulsars 
from the ATNF Pulsar Catalogue with DM measurements 
\citep{Man05}.\footnote{http://www.atnf.csiro.au/research/pulsar/psrcat.}
From Fig. \ref{fig: pulm}, we see that they are primarily distributed in the Galactic disk toward the inner Galaxy.
Different from the statistical analysis by using H$\alpha$ emission, 
as the lines of sight to pulsars pass through both warm and cold phases of the ISM,
the DM SF of pulsars can probe various turbulence characteristics 
in different interstellar phases.

\begin{figure*}[htbp]
\centering   
\subfigure[]{
   \includegraphics[width=8.5cm]{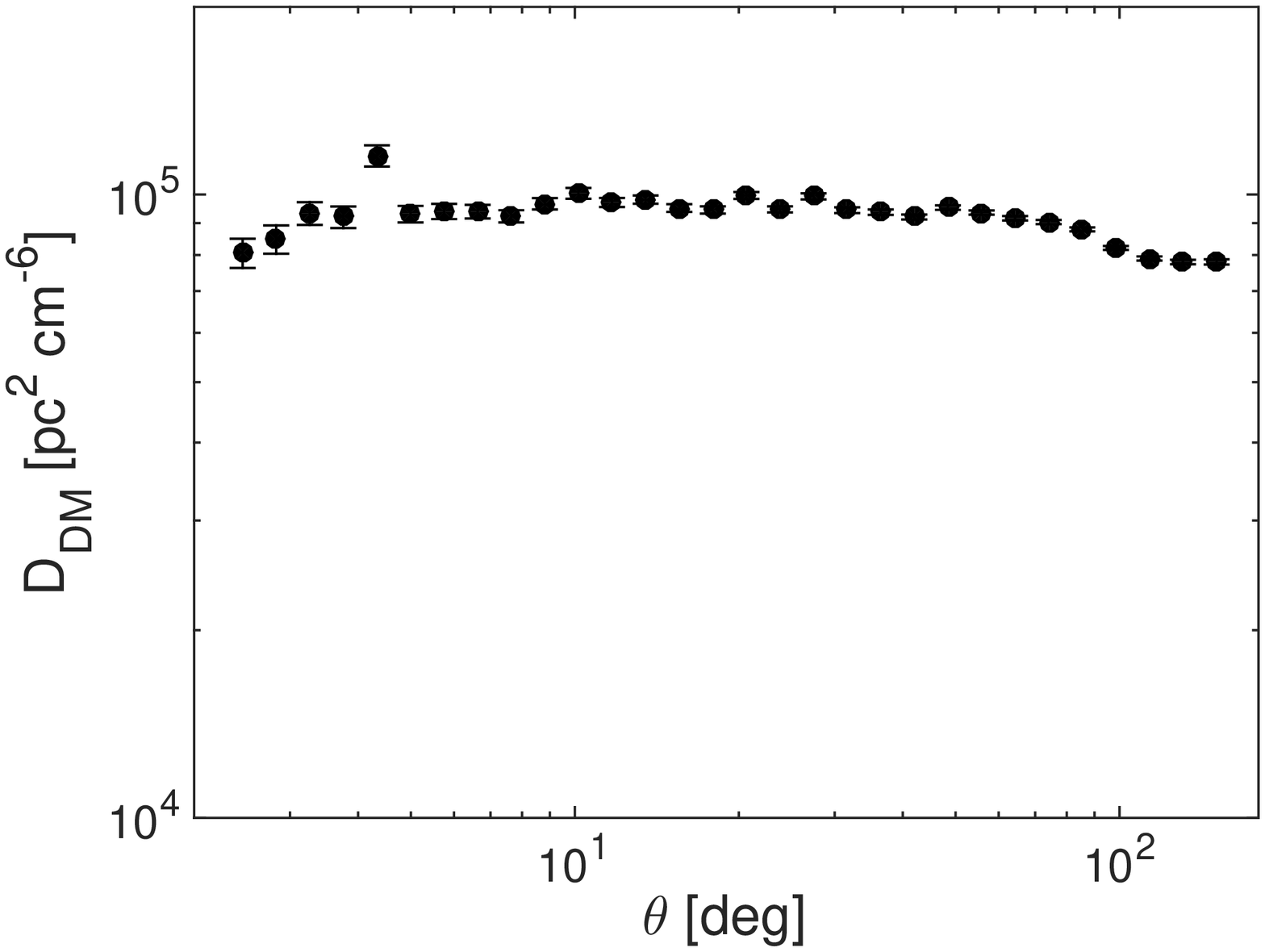}\label{fig: tot}}
\subfigure[]{
   \includegraphics[width=8.5cm]{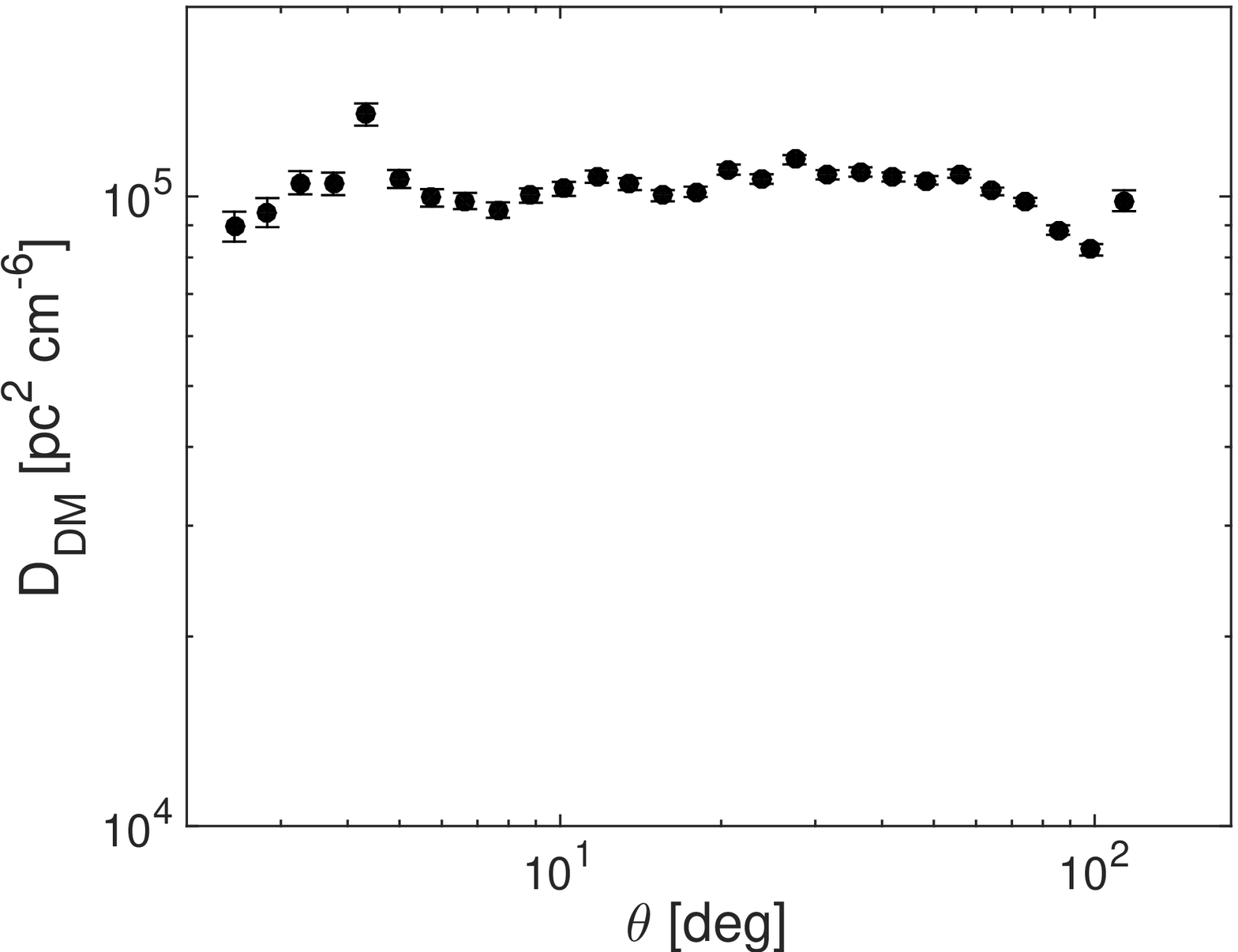}\label{fig: l5par}}
\caption{(a) DM SF for all pulsars.  
(b) DM SF for the pulsars in the inner Galactic disk with $|\text{GB}|<5^\circ$, GL$<60^\circ$, and GL$>300^\circ$. }
\end{figure*}

We first measure the DM SF of all pulsars over the whole sky, 
\begin{equation}
     D_\text{DM}(\theta) = \langle [ \text{DM} (\bm{X_1}) - \text{DM}(\bm{X_2})]^2 \rangle,
\end{equation}
where $\bm{X}$ is the position of a pulsar in the sky,
$\theta$ is the angular separation between pulsars, 
and the average is computed over all pulsar pairs with the same $\theta$.
The angular separations are binned 
evenly on a logarithmic scale within the range $2.5^\circ < \theta <  150.8^\circ$. 
In Fig. \ref{fig: tot}, we see a basically flat DM SF with an insignificant dependence on $\theta$.
It is mainly contributed by the DMs of the pulsars at $|\text{GB}|<5^\circ$ toward the inner Galaxy
(see Fig. \ref{fig: l5par}). 
{Near the midplane of the Galactic disk, 
the measured $\delta n_e$ is too much contaminated by H \uppercase\expandafter{\romannumeral2} regions and spiral arms
\citep{Ock20}, 
and the assumption of statistically homogeneous $n_e$ used in Eq. \eqref{eq: assmfd} is not appropriate.} 
Since the information of interstellar turbulence cannot be effectively extracted from the DMs of very low-latitude pulsars,
we next examine the DM SF of the pulsars at higher latitudes.

(1) The interstellar turbulence in cold phases with a shallow density spectrum ($|\text{GB}|>5^\circ$).

\begin{figure*}[htbp]
\label{fig: gb5gen}
\centering   
\subfigure[]{
   \includegraphics[width=8.5cm]{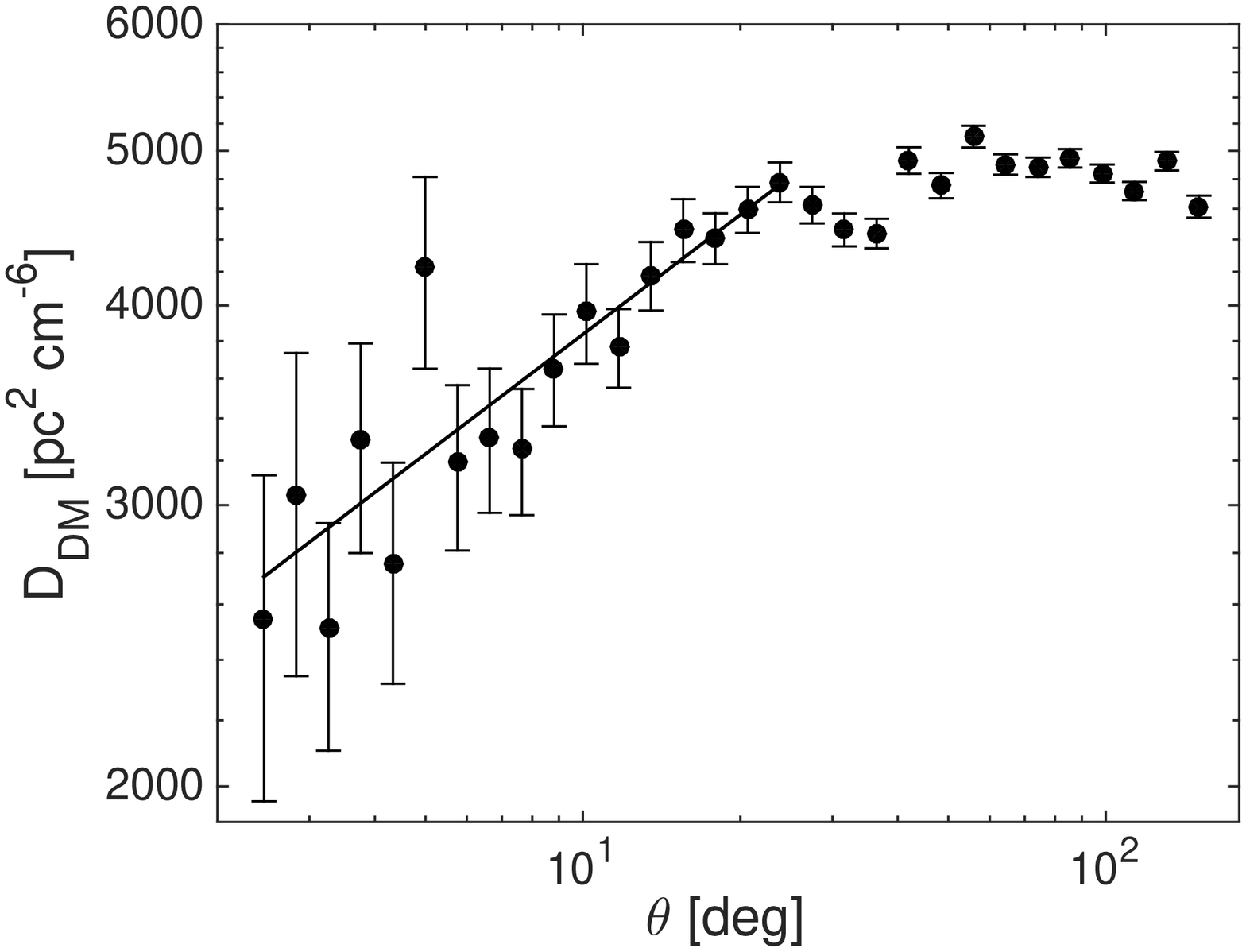}\label{fig: gb5}}
\subfigure[]{
   \includegraphics[width=8.5cm]{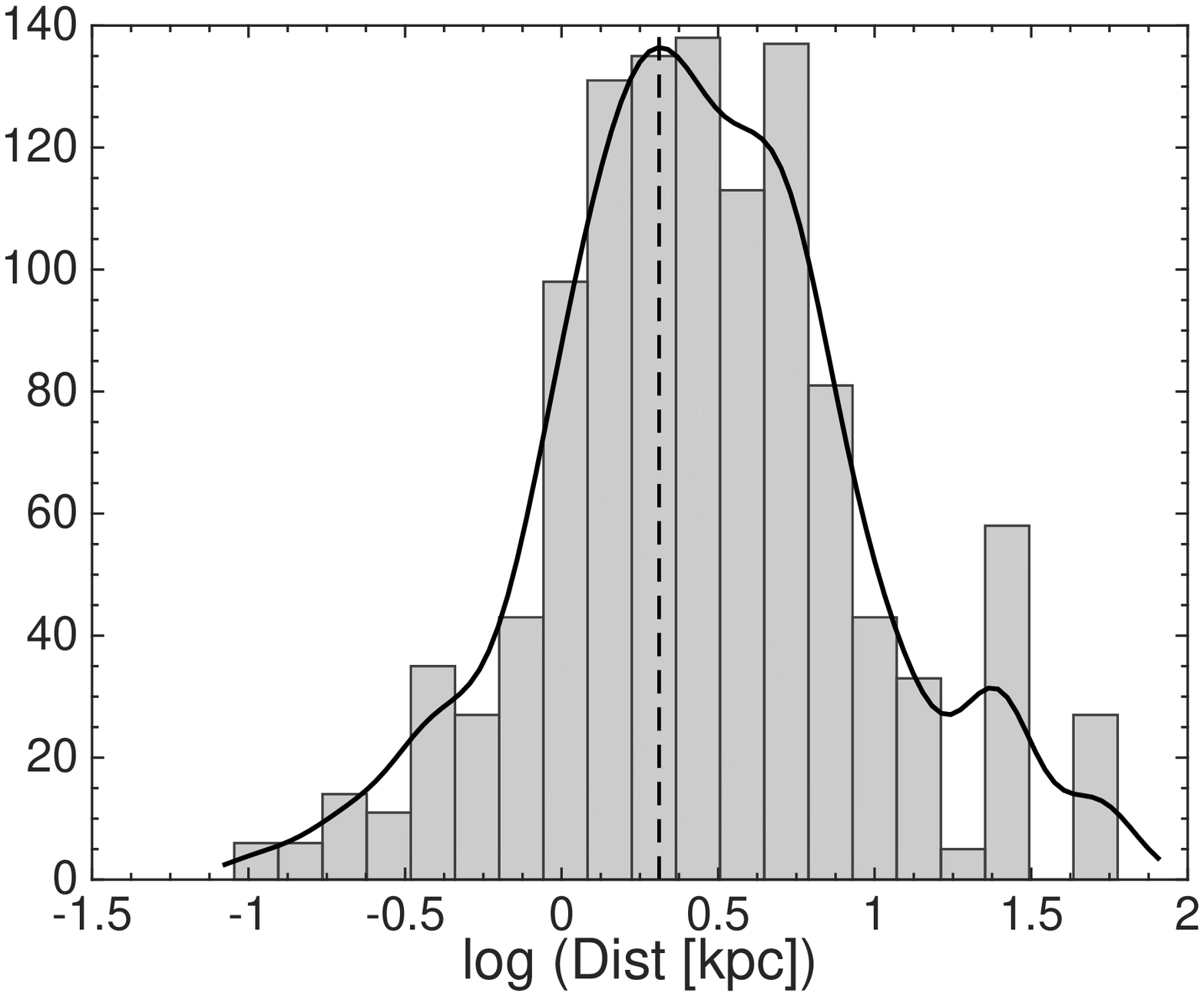}\label{fig: histgb5}}
\caption{ (a) DM SF for $1141$ pulsars at $|\text{GB}|>5^\circ$.
The solid line represents the fit to the data, which is 
described by Eq. \eqref{eq: fitshaw}.
(b) The distribution of the pulsar distances. The kernel density estimation is shown by the thick solid line. 
The peak of the distribution is indicated by the vertical dashed line. }
\end{figure*}

In Fig. \ref{fig: gb5}, we present the DM SF for the pulsars at GB$>5^\circ$ (as indicated as the horizontal solid lines in Fig. \ref{fig: skymap}).
The error bars indicate $95\%$ confidence intervals. 
Here and elsewhere in this paper, all uncertainties are given at $95\%$ confidence.
The fit to the data gives
\begin{equation}\label{eq: fitshaw}
\begin{aligned}
&  D_\text{DM}(2.5^\circ<\theta < 23.8^\circ) [\text{pc}^2  \text{cm}^{-6}] \\
=&    2.2 \times10^3   \pm 3.6\times10^2 (\theta[^\circ])^{0.25 \pm 0.07}.
\end{aligned}
\end{equation}
{We take the value of $\theta$
where $D_\text{DM}$ seems to saturate as the upper bound of the fitting range.}
The fitted power law corresponds to a shallow density spectrum in Case 1 in Section \ref{sec: sfdm}.
To compare with the analytical expression in Eq. \eqref{eq: c1apdts}, 
we adopt $\theta = R/L$ {under the assumption of statistically homogeneous density distribution}. 
We determine the value of $L$ by using the peak of the pulsar distance distribution, i.e., $L\approx 2.0$ kpc (see Fig. \ref{fig: histgb5}).
The pulsar distance Dist is also provided by the ATNF Pulsar Catalogue, 
whose default value is derived based on the YMW16 Galactic electron density model
\citep{Yao17}.
Here and below we only use the distribution of Dist to estimate the value of $L$. 
\footnote{The YMW16 model performs better in estimating pulsar distances, especially for high-latitude pulsars 
\citep{Yao17}, 
than the NE2001 model
\citep{Cor02,Cor03}, but our analysis does not depend on the accuracy of Dist of individual pulsars, 
and thus does not sensitively depend on the Galactic electron density model used for the distance estimation.}

Eq. \eqref{eq: c1apdts} with $R$ replaced by $L \theta$ becomes 
\begin{subnumcases}
     {    D(\theta) \approx \label{eq: coobsh} }
       2  \langle \delta n_e^2 \rangle L_i^m (L+L_i) L^{-m+1} \theta^{-m+1} \nonumber\\
       + \frac{1}{3}   n_{e0}^2 (L-L_i)^2  , 
       ~~~~~~\frac{L_i}{L}\lesssim \theta \lesssim \frac{L+L_i}{2L}, \label{eq: thshalco}\\     
       4  \langle \delta n_e^2 \rangle L_i^m \bigg(\frac{L+L_i}{2}\bigg)^{-m+2} + \frac{1}{3}   n_{e0}^2 (L-L_i)^2  , \nonumber \\
       ~~~~~~~~~~~~~~~~~~~~~~~~~~~~~~~~~~~~~~~~~~~~~~ \theta \gtrsim \frac{L+L_i}{2L}.
\end{subnumcases}
By comparing Eq. \eqref{eq: thshalco} with Eq. \eqref{eq: fitshaw}, one can immediately find
\begin{equation}
     m=0.75,
\end{equation}
which gives $\alpha_{3D} = m -3 = -2.25$ (Eq. \eqref{eq: malp}) for the turbulent density spectrum over 
a range of length scales (i.e., $L\theta$) from $\approx 87$ pc to $831$ pc . 
A shallow density spectrum with $\alpha_{3D}\approx -2.6$ was earlier derived by 
\citet{XuZ17}
from the interstellar scattering measurements of around $100$ high-DM pulsars 
\citep{Kri15}.
\citet{XuZ16} found a similar slope with $\alpha_{3D}\approx-2.64$ of a density spectrum, 
based on the SF of rotation measures of 38 extragalactic sources observed in an area of the sky away from the Galactic plane 
\citep{MS96}, 
which spans over the range $\approx 4$ pc - $100$ pc.
\footnote{\citet{MS96} introduced a different explanation for the slope of the rotation measure SF.}
Here by using a different approach and a large sample of pulsars, 
we found an even shallower density spectrum extending to larger length scales.
In fact, a shallow density spectrum is one of the important characteristics of supersonic turbulence that arises in a cold medium
\citep{BLC05,KL07}.
The excess of small-scale density structures created by supersonic flows 
accounts for the shallow slope of the density spectrum. 
The density spectra obtained by using the tracers of cold phases, 
e.g., CO emission, H\uppercase\expandafter{\romannumeral1} absorption, are usually shallow
(see reviews by \citealt{Laz09rev,HF12}).
As the lines of sight to pulsars pass through the cold phases that inhabit in the Galactic disk,
we naturally expect a shallow density spectrum as seen from their DM SF. 
Compared with earlier measurements, our result suggests the existence of supersonic turbulence on larger length scales. 
A caveat is that the spectral slope can be affected by the non-turbulent density structures along the path, 
leading to a shallower density spectrum that we find from the DMs of pulsars.

The inner scale $L_i$ of the shallow density spectrum is the characteristic size of small-scale density structures, 
which is much less than $L$.
Therefore, we can simplify Eq. \eqref{eq: coobsh} to have 
\begin{subnumcases}
     {    D(\theta) \approx  \label{eq: simc1sht} }
       2  \langle \delta n_e^2 \rangle L_i^m  L^{-m+2} \theta^{-m+1} 
       + \frac{1}{3}   n_{e0}^2 L^2  , \nonumber \\
       ~~~~~~~~~~~~~~~~~~~~~~~~~~~~~~~~~~~~\frac{L_i}{L}\lesssim \theta \lesssim \frac{1}{2}, \\     
       4  \langle \delta n_e^2 \rangle L_i^m \bigg(\frac{L}{2}\bigg)^{-m+2} + \frac{1}{3}   n_{e0}^2 L^2  , \nonumber \\
       ~~~~~~~~~~~~~~~~~~~~~~~~~~~~~~~~~~~~~~~~~~~~~~ \theta \gtrsim \frac{1}{2}. \label{eq: simc1shtb}
\end{subnumcases}
To explain the fit to the observational result in Eq. \eqref{eq: fitshaw}, 
the parameters in Eq. \eqref{eq: simc1sht} should satisfy 
\begin{equation} 
         \bigg(\frac{ \langle \delta n_e^2 \rangle }{ 0.1~\text{cm}^{-6}} \bigg) \bigg(\frac{L_i}{\text{pc}}\bigg)^{0.75} \approx 2.3 ,
\end{equation}
and
\begin{equation}\label{eq: mempd}
      \frac{1}{3}   n_{e0}^2 L^2  <  D_\text{DM} (\theta = 2.5^\circ).
\end{equation}
$L_i$ of a shallow density spectrum on the order of a few pc is also suggested by the rotation measure SF 
\citep{XuZ16}.
As a possible interpretation, 
$L_i$ corresponds to the characteristic size of the density structures {with the typical electron density $\approx 0.3$ cm$^{-3}$}
that undergo the transition from atomic to molecular hydrogen
\citep{Spit78,Elm99}.
The drop of the electron fraction in colder and denser phases results in the cutoff of the density spectrum. 
{A clumpy density distribution with the electron density
on the order of $0.1$ cm$^{-3}$ has also been suggested by earlier studies 
(e.g., \citealt{Cor02,Pet02,Gae08}).}

The condition in Eq. \eqref{eq: mempd} yields
\begin{equation}
      n_{e0}  <  0.046~ \text{cm}^{-3},
\end{equation}
where $L \approx 2.0~$kpc is used. 
{This is consistent with the 
midplane electron density for the thick disk on the order of
$0.01$ cm$^{-3}$ in the YMW16 model
\citep{Yao17}.}
At larger $\theta$, according to Eq. \eqref{eq: simc1shtb}, 
we expect $D_\text{DM}$ to saturate at $\theta \gtrsim 0.5 $ radians, i.e., $\theta \gtrsim28.6^\circ$. 
This agrees with the observational result as shown in Fig. \ref{fig: gb5}.

(2) The interstellar turbulence in the diffuse WIM with a Kolmogorov density spectrum 
($|\text{GB}|>5^\circ$ and $120^\circ<\text{GL}<240^\circ$).

\begin{figure*}[htbp]
\centering   
\subfigure[]{
   \includegraphics[width=8.5cm]{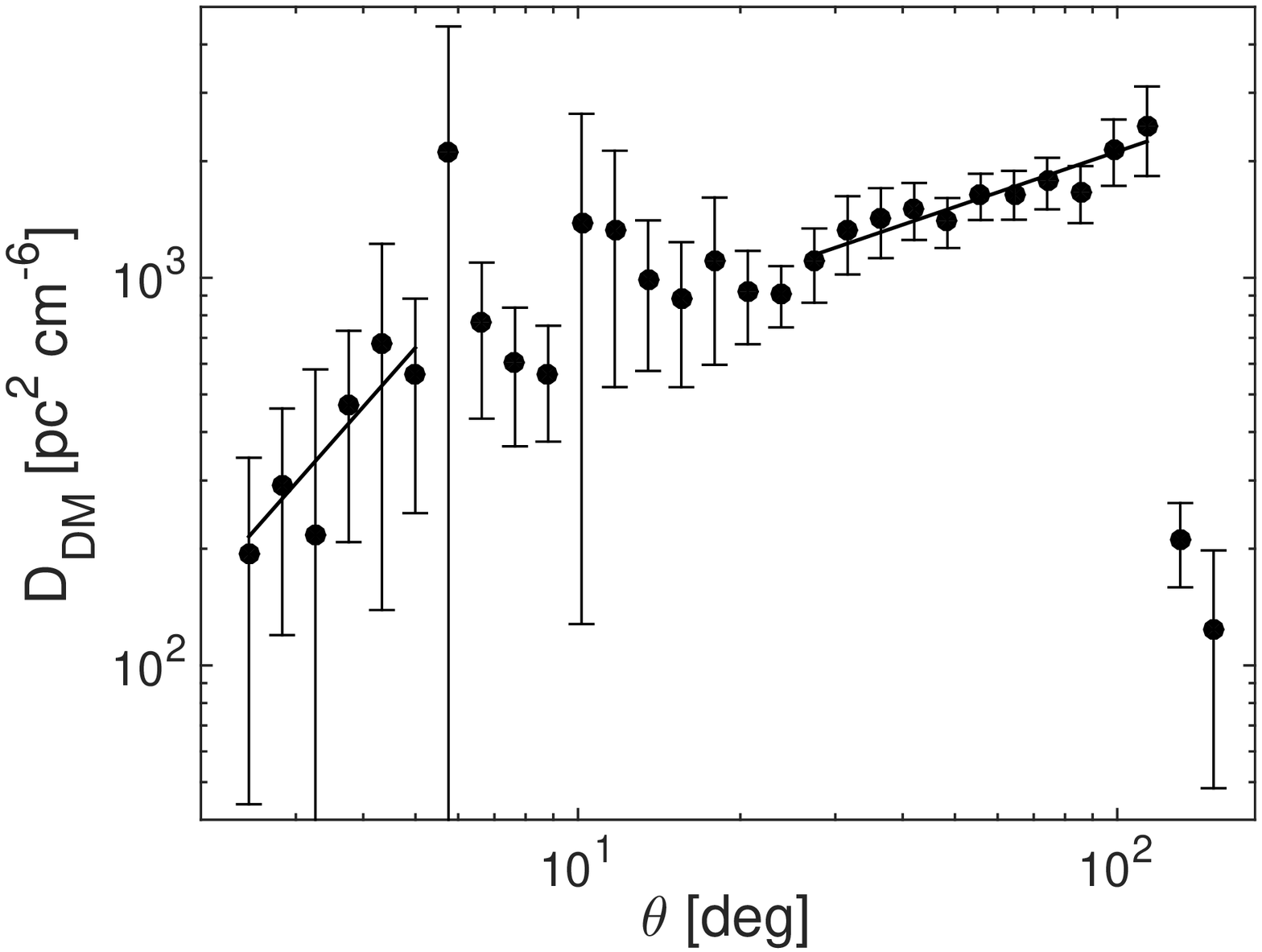}\label{fig: gb5loc}}
\subfigure[]{
   \includegraphics[width=8.5cm]{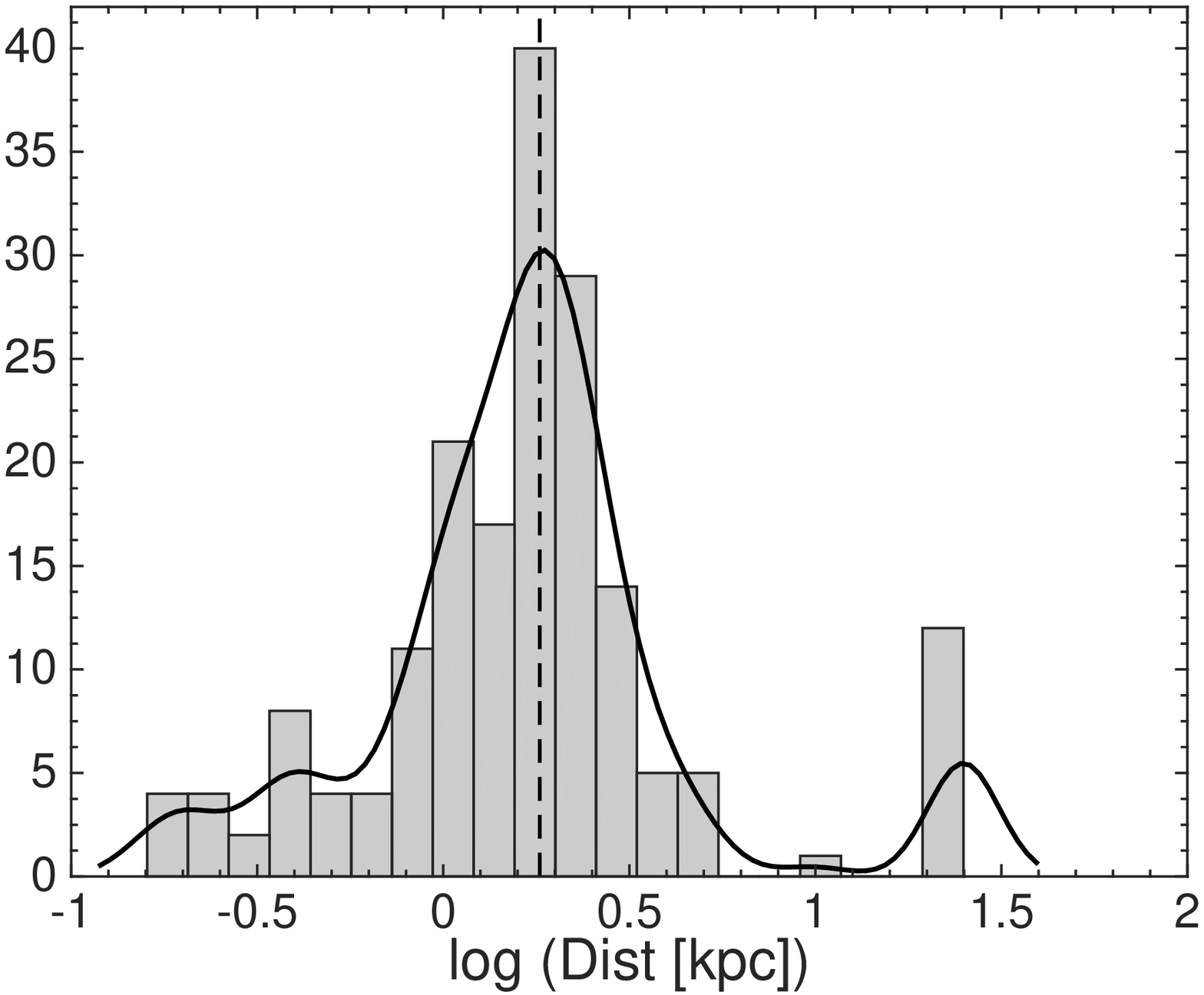}\label{fig: histgb5loc}}
\caption{ Same as Fig. \ref{fig: gb5gen}, but for $181$ pulsars at 
$|\text{GB}|>5^\circ$ and $120^\circ<\text{GL}<240^\circ$. 
The data points at small and large $\theta$ in (a) are 
separately fitted by Eq. \eqref{eq: fipzkost} and Eq. \eqref{eq: fwimlas}, respectively.}
\end{figure*}

To probe the turbulence in the diffuse WIM, we choose the pulsars in an area of the sky with 
$|\text{GB}|>5^\circ$ and $120^\circ<\text{GL}<240^\circ$, which is off the Galactic midplane 
toward the outer Galaxy (see Fig. \ref{fig: skymap}).
Similar to the above analysis
we present their DM SF in Fig. \ref{fig: gb5loc}.
The large error bars come from the small size of the subsample of pulsars used here. 
The fit to the data at small $\theta$ shows 
\begin{equation}\label{eq: fipzkost}
\begin{aligned}
 &  D_\text{DM} (2.5^\circ< \theta <  5.0^\circ) [\text{pc}^2  \text{cm}^{-6}] \\
  =&   52.1\pm 1.1\times10^2 (\theta[^\circ])^{1.58 \pm 1.43}.
\end{aligned}
\end{equation}
The corresponding range of length scales for the fit,
i.e., $L\theta$, is $\approx 79$ pc$-157$ pc, 
where $ L \approx  1.8$ kpc is taken as the peak of the distribution of the pulsar distances (see Fig. \ref{fig: histgb5loc}).
Because of the lower density in the diffuse WIM, 
we see a lower level of DM fluctuations compared with the case in cold interstellar phases (see Eq. \eqref{eq: fitshaw}). 
Based on the fitted power-law index, we find that Eq. \eqref{eq: ca2thess} in Case 2 applies to this situation. 
Under the consideration $R=L\theta$, Eq. \eqref{eq: ca2thess} can be rewritten as 
\begin{subnumcases}
     { D(\theta) \approx   }
       2  \langle \delta n_e^2 \rangle L_i^{-m} (L+L_i) L^{m+1}\theta^{m+1}   \nonumber\\
       + \frac{1}{3} \big(\langle \delta n_e^2 \rangle + n_{e0}^2 \big) (L-L_i)^2 , ~~~~~~~~~\theta \lesssim \frac{L_i}{L},\\    
       2  \langle \delta n_e^2 \rangle L_i (L+L_i) + \frac{1}{3} \big(\langle \delta n_e^2 \rangle +n_{e0}^2\big)  (L-L_i)^2  , \nonumber \\
       ~~~~~~~~~~~~~~~~~~~~~~~~~~~~~~~~~~~~~~~~~~~~~~~~~~~~~~~~\theta \gtrsim \frac{L_i}{L}.
\end{subnumcases}
For a steep density spectrum, $L_i $ coincides with the energy injection scale of turbulence.
It is believed to be in the range $\sim 50-500~$ pc in the ISM   
\citep{ElmegreenScalo,Chep10,Bec16}, 
with the main driver as supernova explosions
\citep{Pad16}.
As $L_i$ is much less than $L$, 
we recast the above equation into 
\begin{subnumcases}
     { D(\theta) \approx   }
       2  \langle \delta n_e^2 \rangle L_i^{-m}  L^{m+2}\theta^{m+1}   \nonumber\\
       + \frac{1}{3} \big(\langle \delta n_e^2 \rangle + n_{e0}^2 \big) L^2 , ~~~~~~~~~~~~\theta \lesssim \frac{L_i}{L}, \label{eq: psikolsth}\\    
       2  \langle \delta n_e^2 \rangle L_i L + \frac{1}{3} \big(\langle \delta n_e^2 \rangle +n_{e0}^2\big)  L^2  , \nonumber \\
       ~~~~~~~~~~~~~~~~~~~~~~~~~~~~~~~~~~~~~~~~~~~~~~~~\theta \gtrsim \frac{L_i}{L}.
\end{subnumcases}
The comparison between Eq. \eqref{eq: psikolsth} and Eq. \eqref{eq: fipzkost}
provides constraints on the parameters, 
\begin{subequations}
\begin{align}
   &   m  = 0.58,\\
   & \Big(\frac{\langle \delta n_e^2 \rangle}{ 2\times 10^{-4}~ \text{cm}^{-6}}\Big) \Big(\frac{L_i}{ 100 ~\text{pc}}\Big)^{-0.58} \approx 4.5,\\
   &   \frac{1}{3} \big(\langle \delta n_e^2 \rangle + n_{e0}^2 \big)L^2  < D_\text{DM} (\theta = 2.5^\circ), \label{eq: tadekolu}
\end{align}
\end{subequations}
where $L \approx  1.8$ kpc is used. 
Here Eq. \eqref{eq: tadekolu} gives 
\begin{equation}
     \sqrt{\langle \delta n_e^2 \rangle + n_{e0}^2}< 0.014 ~\text{cm}^{-3}.
\end{equation}
The resulting spectral index 
$\alpha_\text{3D} = -m -3 = -3.58$ (Eq. \eqref{eq: malp})
is very close to the Kolmogorov slope. This is expected for the turbulence in the diffuse WIM 
\citep{Armstrong95,CheL10}.
Since the diffuse WIM is not preferentially populated by pulsars, the limited size of the
sample {and the contaminations from, e.g., H \uppercase\expandafter{\romannumeral2} regions, along the path} 
prevent us from obtaining an extended density spectrum 
over many orders of magnitude in scales 
as the {composite spectrum 
obtained from the scintillations and scattering of nearby pulsars 
in the local ISM
\citep{Armstrong95}.}

As shown in Fig. \ref{fig: gb5loc}, instead of saturating at large $\theta$, 
the SF exhibits another power law, which will be discussed below.

(3) Galactic-scale turbulence with a steep density spectrum ($|\text{GB}| > 10^\circ$).

\begin{figure*}[htbp]
\centering   
\subfigure[]{
   \includegraphics[width=8.5cm]{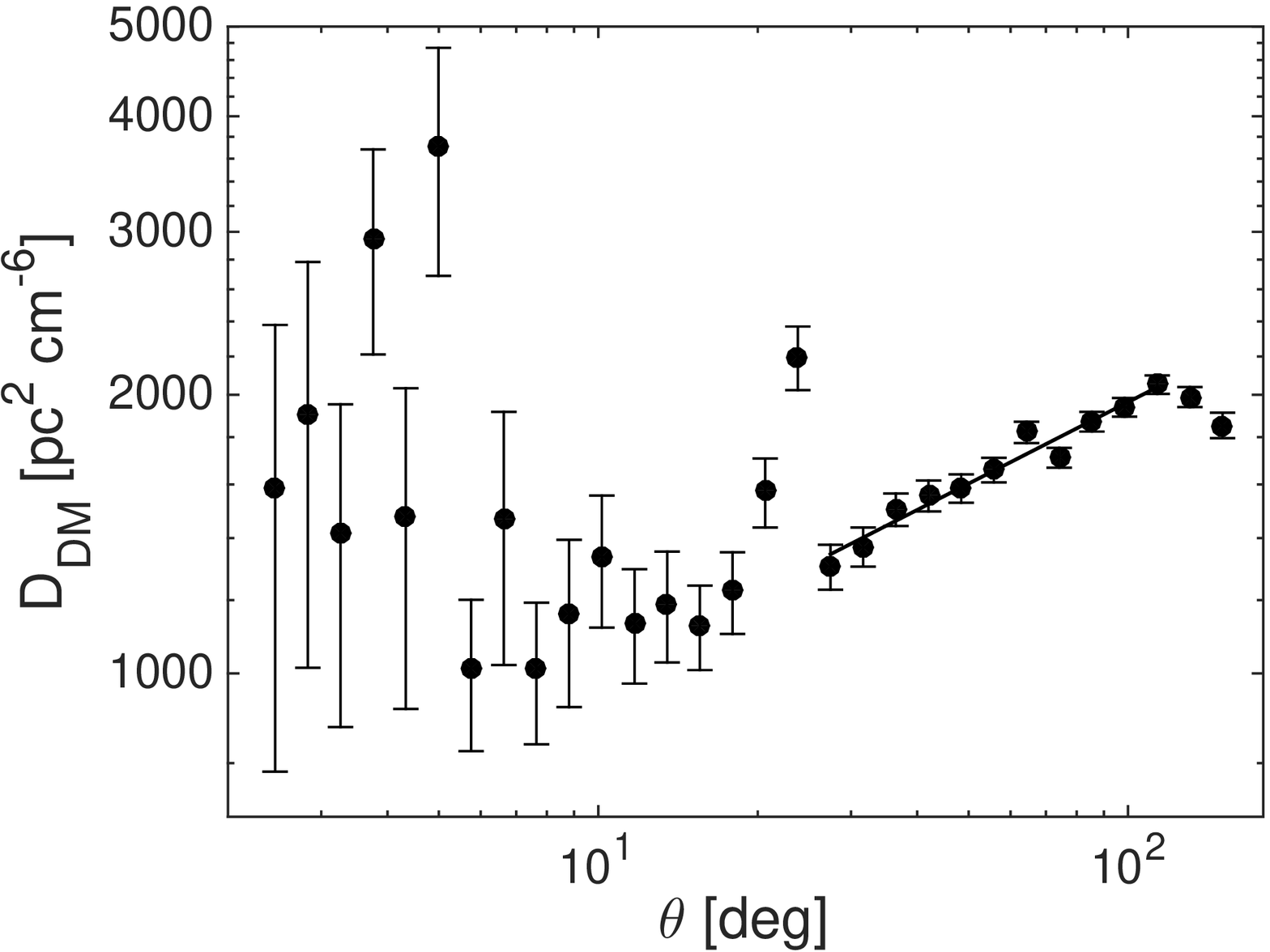}\label{fig: gb10}}
\subfigure[]{
   \includegraphics[width=8.5cm]{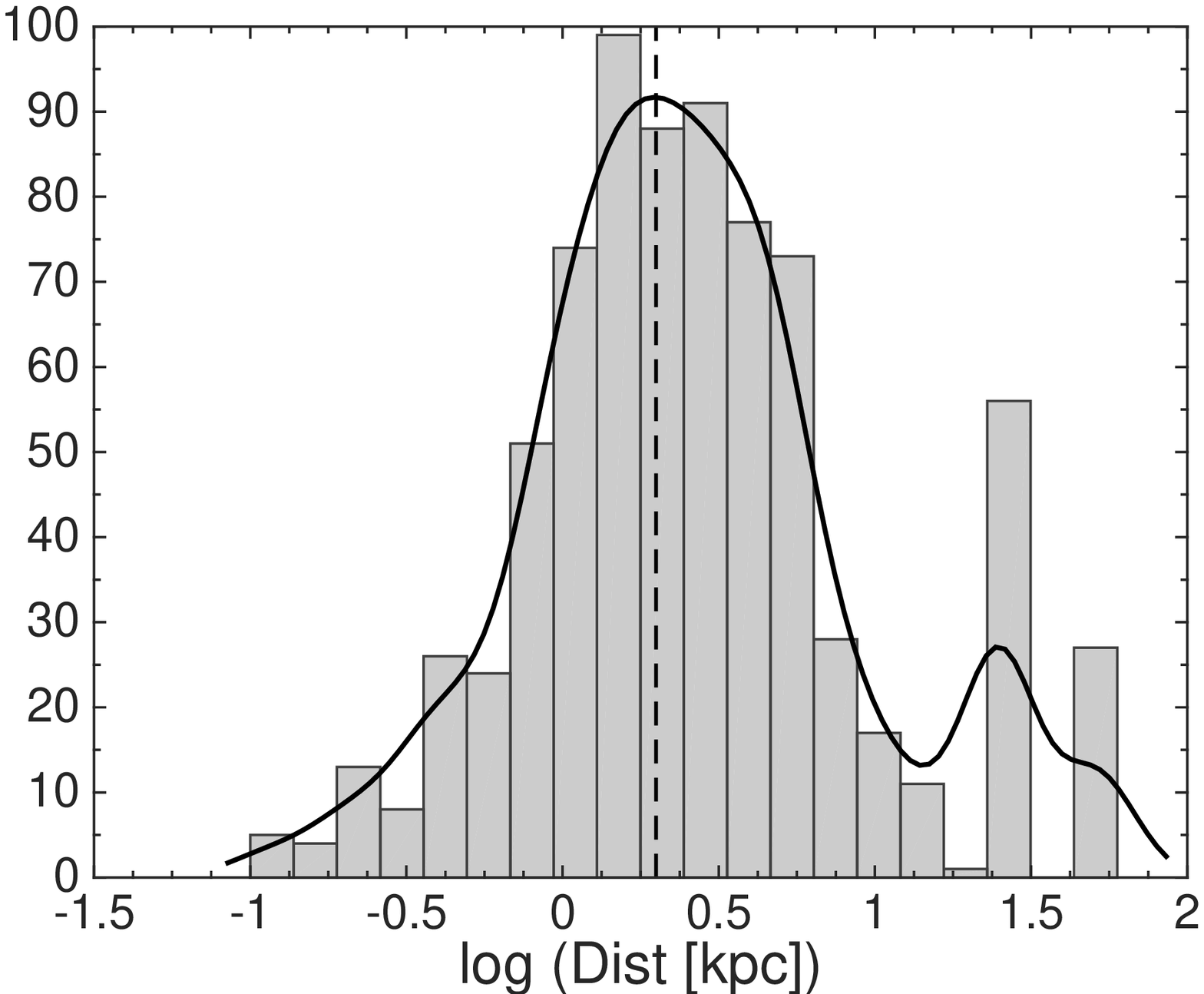}\label{fig: histgb10}}
\caption{ Same as Fig. \ref{fig: gb5gen}, but for $773$ pulsars at $|\text{GB}| > 10^\circ$.
The fit in (a) is given by Eq. \eqref{eq: slaskls}. }
\end{figure*}

In the above analysis, we find a separate power law at large $\theta$ in the WIM, which can be fitted by 
\begin{equation}\label{eq: fwimlas}
\begin{aligned}
   & D_\text{DM} (27.4^\circ<\theta < 113.5^\circ) [\text{pc}^2 \text{cm}^{-6}] \\
     = & (2.4 \pm 1.4) \times10^2  (\theta[^\circ])^{0.47\pm 0.14}.
\end{aligned}
\end{equation}
A similar power law can also be found for a larger sample of pulsars 
that lie over the entire range of longitude but at 
higher latitudes. 
Fig. \ref{fig: gb10} displays the DM SF for the pulsars at $|\text{GB}| > 10^\circ$. 
The SF at small $\theta$ is dominated by non-turbulent density structures {in the inner Galaxy.} 
At large $\theta$, it can be fitted by  
\begin{equation}\label{eq: slaskls}
\begin{aligned}
   &  D_\text{DM} ( 27.4^\circ < \theta < 113.5^\circ ) [\text{pc}^2  \text{cm}^{-6}] \\
         = &( 5.1 \pm  1.0) \times10^2  (\theta[^\circ])^{0.29 \pm 0.05}.
\end{aligned}
\end{equation}
We assume that the power-law behavior of the SF at large $\theta$ is also of turbulent origin. 
Based on the SF slope and the range of $\theta$, we find that it can be approximately treated as Case 3 in Section \ref{sec: sfdm}. 

We rewrite Eq. \eqref{eq: ca3sthth} by replacing $R$ with $L\theta$,
\begin{subnumcases}
     { D(\theta)  \approx  }
       2  \langle \delta n_e^2 \rangle L_i^{-m} L^{m+2} \theta^{m+1} + \frac{1}{3} \big(\langle \delta n_e^2 \rangle + n_{e0}^2 \big) L^2, \nonumber \\
     ~~~~~~~~~~~~~~~~~~~~~~~~~~~~~~~~~~~~~~~~~~~~~~~~~~~~~~~~  \theta \lesssim \frac{1}{2}, \label{eq: segsmal} \\   
         \langle \delta n_e^2 \rangle  L_i^{-m} L^{m+2} \theta^m + \frac{1}{3} \big(\langle \delta n_e^2 \rangle + n_{e0}^2\big) L^2, \nonumber \\
        ~~~~~~~~~~~~~~~~~~~~~~~~~~~~~~~~~~~~~~~~~~~~~~ \frac{1}{2} \lesssim \theta \lesssim \frac{L_i}{L},  \label{eq: fimedco}\\      
         \langle \delta n_e^2 \rangle L^2 +  \frac{1}{3} \big(\langle \delta n_e^2 \rangle + n_{e0}^2 \big) L^2, ~~~ \theta \gtrsim \frac{L_i}{L}. 
\end{subnumcases}
We adopt the fit in Eq. \eqref{eq: slaskls}, as it is obtained from a larger sample size compared with Eq. \eqref{eq: fwimlas}. 
A comparison between Eq. \eqref{eq: slaskls} and 
Eq. \eqref{eq: fimedco} yields 
\begin{subequations}
\begin{align}
     &  m  = 0.29,  \\
     & \Big(\frac{\langle \delta n_e^2 \rangle}{10^{-4}~ \text{cm}^{-6}}\Big) \Big(\frac{L_i}{2~ \text{kpc}}\Big)^{-0.29} \approx 4.1,\\
     &  \frac{1}{3}\big( \langle \delta n_e^2 \rangle + n_{e0}^2 \big) L^2 < D_\text{DM} (\theta = 27.4^\circ),   \label{eq: ladecon}
\end{align}
\end{subequations}
where Eq. \eqref{eq: ladecon} implies 
\begin{equation}
    \sqrt{ \langle \delta n_e^2 \rangle+n_{e0}^2} < 0.032 ~\text{cm}^{-3}.
\end{equation}
Here we use $L \approx 2.0~$kpc as derived from the distribution of the pulsar distances (see Fig. \ref{fig: histgb10}).
We note that the driving scale of turbulence $L_i$ is larger than $L$ (see Case 3 in Section \ref{sec: sfdm}),
but appears to be of the same order of magnitude as $L$.
This is consistent with the length scale $\approx 4.0$ kpc corresponding to $113.5^\circ$ where the SF reaches its maximum value. 
Over the range of length scales $\approx 956$ pc$- 4.0$ kpc, 
{which extends to scales larger than the 
scale height $\approx 1.7$ kpc of the thick disk in the YMW16 model
\citep{Yao17},}
the 1D power-law index of the density spectrum is 
$\alpha_\text{1D} = -m -1 = - 1.29$.
It is a bit shallower than the Kolmogorov slope ($-5/3$) as expected for 2D turbulence in the inverse cascade regime 
in a nearly incompressible medium
\citep{Kraid67,Das05}.
The expected steepening of SF 
at $\theta \lesssim 0.5$ radians, i.e., $\theta \lesssim 28.6^\circ$, in Eq. \eqref{eq: segsmal} 
is not seen in Fig. \ref{fig: gb10} due to the presence of non-turbulent density structures on smaller length scales.

The above Galactic-scale density spectrum 
can only been seen in the diffuse WIM.
Its effect on shaping the density structures near the Galactic plane 
is insignificant (see Fig. \ref{fig: gb5}). 
A flat Galactic magnetic energy spectrum with a 1D power-law index $ - 0.37$
over the range $0.5-15$ kpc 
was earlier measured by 
\citet{Han04},
based on the rotation and dispersion measures of pulsars. 
For external spiral galaxies, density and velocity power spectra extending to scales comparable to the size of galactic disk 
have also been reported
\citep{Dut13,Nan20}. 
Gravitational instabilities are believed to be the main driver of such galactic-scale turbulence, 
especially for the galaxies with relatively high star formation rates 
\citep{Elm03,Rom14,Krum16}.
An inverse cascade in a quasi-2D disk can also contribute to the large-scale turbulence
\citep{Wad02,Bour10}.

\section{Discussion and conclusions}

We summarize the density spectra obtained from the DM SF of pulsars 
in Table \ref{tab:com}.
They reveal different properties of turbulence
depending on the interstellar phase and the range of length scales. 
Our results demonstrate that the DMs of pulsars bear unique signatures of interstellar turbulence and can be used 
synergistically with other statistical techniques to 
achieve a comprehensive picture of interstellar density structures and distribution.

Earlier models for the Galactic electron density distribution, 
e.g., 
the NE2001 model
\citep{Cor02,Cor03},
the YMW16 model
\citep{Yao17},
include multiple components and involve a large number of parameters
to account for the basic structure with thin and thick disks, spiral arms, and other local features.
Differently, here we focus on the statistics of electron density fluctuations ($\delta n_e$ in Eq. \eqref{eq: assmfd}), 
which are only characterized by turbulence parameters and 
can be used to statistically explain the observations of a large number of pulsars
\citep{XuZ17}
and other turbulence-related observations
\citep{XuZ16}.

\begin{table*}[!htbp]
\renewcommand\arraystretch{1.5}
\centering
\begin{threeparttable}
\caption[]{Electron density spectra in the ISM obtained from DMs of pulsars}\label{tab:com} 
  \begin{tabular}{|c|c|c|c|c|c|}
     \toprule
       Area of sky                 &   Interstellar phase     &        Measured range of length scales        &   $\sqrt{\langle \delta n_e^2 \rangle}$  [cm$^{-3}$]  &    $L_i$    &    $\alpha_\text{1D}$      \\
      \hline
     $|\text{GB}|>5^\circ$   &  Cold phases     &       $ 87$ pc$-831$ pc            & $\sim 0.3$  & $\sim$ pc  (inner scale)    &   $-0.25$  \\
     $|\text{GB}|>5^\circ$ and $120^\circ<\text{GL}<240^\circ$ & WIM & $ 79$ pc$-157$ pc & $\sim 0.01$  & $\sim100$ pc (outer scale) & $ -1.58$ \\
     $|\text{GB}| > 10^\circ$   &   WIM             &       $ 956$ pc$- 4.0$ kpc        & $\sim 0.01$ &  $ 4.0$ kpc (outer scale)   & $ - 1.29$  \\
     \bottomrule
    \end{tabular}
 \end{threeparttable}
\end{table*}

Compared with the rotation measure SF and scatter broadening of pulsars, 
without contributions from magnetic fields or invoking the scattering effect, 
the DM SF of pulsars provides a relatively clean and direct measurement of density fluctuations over a range of length scales. 
Here we found a shallow density spectrum {extending to larger scales compared with our earlier studies in
\citet{XuZ16} and \citet{XuZ17}.} 
Its inner cutoff scale on the order of pc, which was also suggested by the rotation measure SF
\citep{XuZ16},
might be related to the phase transition and formation of molecular hydrogen. 
Within molecular clouds, 
the interstellar scattering is more suitable than the DM SF of pulsars to probe the sub-pc density fluctuations
\citep{XuZ17}.
As shallow density spectra have also been measured with cold gas tracers both at high latitudes 
(e.g., \citealt{Chep10})
and in molecular clouds 
\citep{HF12}, 
our methods can be used synergistically with cold gas tracers to probe 
the supersonic turbulence in cold interstellar phases.

Compared with the statistical measurement of density fluctuations using the diffuse H$\alpha$ emission, 
as pulsars are primarily distributed in the Galactic disk, they can be used to probe the cold gas phases at low latitudes (see above). 
However, the DMs of pulsars are sensitive to 
all density fluctuations along the lines of sight and thus are more subjected to non-turbulent ``noises" than 
the high-latitude H$\alpha$ emission in measuring turbulent density spectra. 
It means that the DMs of pulsars are less favored than the diffuse H$\alpha$ emission to study the turbulence in the diffuse WIM. 
In addition, different from the statistical measurements using spatially continuous emission, when we use 
discrete point sources,
the angular resolution is determined by the projected separation between two point sources and their distances from the observer. 
It determines the minimum length scale for the SF analysis of turbulent fluctuations. 
We note that the Kolmogorov density spectrum seen from the diffuse H$\alpha$ emission was also 
inferred from radio scattering observations of nearby pulsars on scales down to $10^6$ m
\citep{Armstrong95}, 
but this is only for the measurement of the local ISM,
where the turbulence is not supersonic and the lines of sight do not pass through the  
non-turbulent density structures in the disk.

Besides the shallow density spectrum in cold phases and the Kolmogorov-like density spectrum in the diffuse WIM,
we also found a steep density spectrum on very large scales, 
suggesting the existence of galactic-scale turbulence in the Milky Way similar to other external spiral galaxies. 
Its interaction with the smaller-scale interstellar turbulence and its role in e.g., gas dynamics and 
star formation deserves further investigation.

It is important to stress that different from 
the statistical measurements of turbulent velocities, 
density statistics do not directly show the dynamics and energy cascade of turbulence.
They preferentially trace the compressive component of turbulence, which is responsible for generating large density contrasts. 
Therefore, the measured shallow density spectrum should not be treated as the evidence for the 
dominance of compressive component over
solenoidal component of turbulence in the disk. 
The latter, in fact, can be dominant across most of the volume in cold cloud phases 
\citep{Pad16,Qi18,Xu20}.

\acknowledgments

S.X. acknowledges the support for 
this work provided by NASA through the NASA Hubble Fellowship grant \# HST-HF2-51473.001-A awarded by the Space Telescope Science Institute, which is operated by the Association of Universities for Research in Astronomy, Incorporated, under NASA contract NAS5- 26555.

The Wisconsin H-Alpha Mapper and its Sky Survey have been funded
primarily through awards from the U.S. National Science Foundation.

\bibliographystyle{apj.bst}
\bibliography{xu}

\end{document}